# Exploring the magnetic properties of individual barcode nanowires using wide-field diamond microscopy

Jungbae Yoon[1,+], Jun Hwan Moon[2,+], Jugyeong Jeong[1], Yu Jin Kim[3], Kihwan Kim[1], Hee Seong Kang[4], Yoo Sang Jeon[5], Eunsoo Oh[2], Sun Hwa Lee[6], Kihoon Han[7,8], Dongmin Lee[7,9], Chul-Ho Lee[4], Young Keun Kim[2,3*], and Donghun Lee[1,*]

[1]Department of Physics, Korea University, Seoul, Republic of Korea

[2]Department of Materials Science and Engineering, Korea University, Seoul, Republic of Korea

[3]Institute for High Technology Materials and Devices, Korea University, Seoul, Republic of Korea

[4]KU-KIST Graduate School of Converging Science and Technology, Korea University, Seoul, Republic of Korea

[5]Center for Hydrogen·Fuel Cell Research, Korea Institute of Science and Technology, Seoul, Republic of Korea

[6]Center for Multidimensional Carbon Materials (CMCM), Institute for Basic Science (IBS), Ulsan 44919, Republic of Korea

[7]BK21 Graduate Program, Department of Biomedical Sciences, Korea University College of Medicine, Seoul, Republic of Korea

[8]Department of Neuroscience, Korea University College of Medicine, Seoul, Republic of Korea

[9]Department of Anatomy, Korea University College of Medicine, Seoul, Republic of Korea

[+] These authors contributed equally to this work.

[*]**Corresponding authors:**
Email addresses: ykim97@korea.ac.kr (Y. K. Kim) and donghun@korea.ac.kr (D. Lee)

**Abstract**

Barcode magnetic nanowires typically comprise a multilayer magnetic structure in a single body with more than one segment type. Interestingly, owing to selective functionalization and novel interactions between the layers, barcode magnetic nanowires have attracted significant attention, particularly in the field of bioengineering. However, an analysis of their magnetic properties at the individual nanowire level remains challenging. With this background, herein, we investigated the characterization of magnetic nanowires at room temperature under ambient conditions based on magnetic images obtained via wide-field quantum microscopy with nitrogen-vacancy centers in diamond. Consequently, we could extract critical magnetic properties, such as the saturation magnetization and coercivity, of single nanowires by comparing the experimental results with those of micromagnetic simulations. This study opens up the possibility for a versatile characterization method suited to individual magnetic nanowires.

**Main**

In solid-state physics, understanding the magnetic characteristics of low-dimensional nanostructures is considerably interesting. In such structures, the reduced dimensionalities often result in the development of new properties or functionalities that are extremely different from those of their bulk counterparts[1-5]. In addition to this fundamental interest, one-dimensional magnetic nanostructures, such as nanowires and nanotubes, have been widely investigated in the fields of nanotechnology and bioengineering owing to their potential use as logic units, data memory units, biomedical sensors, and drug delivery robots[6-12]. Consequently, extensive research has been conducted to understand the characteristic properties of magnetic nanowires depending on their structural shapes, dimensions, and chemical compositions[13-15]. Recently, this interest has extended further to multi-component systems, such as barcode magnetic nanowires (BMNs), which consist of more than one segment type placed in an alternating fashion in a single body[16-19].

Owing to the selective functionalization and induced interactions between adjacent materials, BMNs have emerged as novel platforms, particularly for bioengineering applications. For instance, a sensitive immunoassay platform has been introduced through the selective conjugation of different antibodies to magnetic and nonmagnetic layers in BMNs[10]. Accordingly, controlled adhesion and differentiation of stem cells and macrophages have been demonstrated by tuning the periodicity and sequence of arginine–glycine–aspartic acid (RGD)–bearing and RGD-free layers in BMNs[20,21]. Freestyle swimming nano-robots have been developed by attaching magnetic and nonmagnetic nanowires using a flexible material sandwiched between them[22].

Despite the notable progress in the growth, fabrication, and control of magnetic nanowires, detailed analysis and microscopic imaging of individual nanowires remain challenging. Magnetometry tools, such as vibrating sample magnetometers (VSMs) or alternating gradient magnetometers, have been employed to analyze the properties of an ensemble (or array) of nanowires. However, the magnetic behavior of individual nanowires can differ from that of the ensemble owing to the magnetostatic interactions between the nanowires[23]. Moreover, the direct imaging of internal structures, magnetic domains, and different material segments within a single nanowire requires magnetic imaging with a high spatial resolution. Several different types of sensitive magnetic measurements or scanning magnetometry experiments have been conducted for the quantitative analysis or high-resolution imaging of magnetic nanowires. For instance, dynamic cantilever magnetometry and superconducting quantum interference devices have been used to obtain the magnetic hysteresis curves of single nanowires and nanotubes, revealing a bistable magnetic behavior or reversal mechanism[1,2,24].

Moreover, magnetic force microscopy, magneto-optical Kerr effect microscopy, X-ray magnetic circular dichroism microscopy, and Lorentz transmission electron microscopy have successfully been applied for imaging internal magnetic structures based on the contrast in the stray field or differences in the X-ray absorption or the electron beam phase[25-27]. Nonetheless, the majority of these experiments require high vacuum and cryogenic environments and they usually do not offer quantitative magnetic information.

In this study, we demonstrate the characterization and microscopic imaging of individual magnetic nanowires and BMNs at room temperature under ambient conditions using wide-field quantum microscopy based on diamond nitrogen-vacancy (NV) centers. Diamond NV centers are solid-state spin qubits that are highly sensitive to the local magnetic field, and the sensing results are well quantified without further calibrations[28-30]. They can also maintain sensing capabilities over a wide range of temperatures, including room temperature, which is an essential requirement for bioengineering applications based on magnetic nanowires. These unique properties indicate the potential of NV centers in the study of magnetic nanowires. In our analysis, to obtain spatially resolved magnetic images, we combined the sensing experiment of NV centers with wide-field-of-view optical microscopy[31]. The employed microscope is relatively easy to operate compared with scanning probe-type magnetometers. Moreover, it provides large images (e.g., >100 µm) with a spatial resolution of several hundred nanometers, enabling the study of multiple nanowires from a single imaging experiment.

Using wide-field diamond microscopy, we mapped the stray field around individual magnetic nanowires, for example, Fe and Co, and extracted their magnetic properties, such as the saturation magnetization ($M_s$) and coercivity ($H_c$), based on a comparison of the data with micromagnetic simulation results. We repeated the measurements at various external magnetic fields and obtained magnetic hysteresis curves for each nanowire. The results revealed magnetically harder ferromagnetic behaviors than their bulk counterparts, agreeing well with the results of previous studies[1,2]. We also studied various BMNs comprising different materials and layer sequences. For instance, we compared BMNs with alternating layers of nonmagnetic and ferromagnetic materials, for example, Au–Fe, or two different ferromagnetic materials, for example, Co–Fe. Note that these BMNs are indistinguishable based on their optical images; however, the magnetic images reveal distinct stray field profiles depending on the material compositions and layer sequences. This enabled us to identify different BMNs and characterize each layer of the material within the BMNs. This study provides an ambient, easy-to-use quantitative analysis method for individual magnetic nanowires and BMNs.

Figure 1(a) illustrates schematics of the wide-field diamond microscope and the measurement principle of NV centers. We placed magnetic nanowires on top of a diamond plate ($2 \times 2 \times 0.1$ mm$^3$) containing an ensemble of NV centers at a concentration of approximately 10 ppm. The NV centers were uniformly distributed over the plate at a constant depth of approximately 15 nm from the surface. We designed an optical microscope to illuminate the NV centers using a green laser ($\lambda = 532$ nm; power = 500 mW) and to record their fluorescence within the field-of-view, for example, ~100 µm. We used a complementary metal–oxide–semiconductor (CMOS) camera to capture the fluorescence images. To avoid unwanted heating of the nanowires by the high-power laser beam, we adopted the total internal reflection fluorescence (TIRF) configuration. A negatively charged NV center exhibits S = 1 triplet states at the ground level. The photoluminescence (PL) signal of NV centers strongly depends on the spin states, enabling the measurement of optically detected magnetic resonance (ODMR)[28,29]. The omega shape of a gold strip line fabricated on a slide cover glass was placed underneath the diamond plate, and it provided microwave radiation at ~2.9 GHz, which was necessary for the transition between spin states. Moreover, in the presence of a nonzero magnetic field along the crystal axis of NV centers, the degenerated $m_s = +1$ and $m_s = -1$ spin states are separated from each other via the Zeeman effect. Accordingly, we probed the field component along the axis direction by measuring the amount of Zeeman shift, as depicted in the ODMR measurement in Fig. 1(a). Owing to the existence of four possible crystal axes of the NV centers in diamond, our measurement based on the NV ensemble provided field information along four different directions, enabling vector magnetometry. We synthesized magnetic nanowires and BMNs using the electrodeposition method with porous anodized aluminum oxide (AAO) templates. We examined the morphology of the nanowires using scanning electron microscopy (SEM). More information regarding the experimental setup, measurement principle, material synthesis, and characterization is summarized in the Methods section and Supplementary Materials.

Figure 1(b) presents examples of the magnetic images of an Fe nanowire obtained along two different axes of the NV centers, i.e., $NV_1 \parallel [111]$ and $NV_3 \parallel [\bar{1}\bar{1}1]$, where the in-plane direction of the former NV center is nearly parallel to the long axis of the nanowire, whereas that of the latter NV center is parallel to the short axis (the images with the other two axes are presented in Fig. S3). The magnetic images evidently present distinct characteristics designated by dipole-like features at the tips of the nanowires. Because the stray field is the largest and enters in and out at the tips, the field projected onto the NV axis appears as a dipole with a direction parallel to the NV axis. This is illustrated in the simulated images in Fig. 1(b).

The simulation consisted of three main components: production of a three-dimensional field around a magnetic nanowire, analysis of the field projected onto an NV axis, and convolution of the ODMR data with a point spread function for single emitters. First, we obtained the magnetic field distribution around a nanowire using an open-source software called the object oriented micromagnetic framework (OOMMF)[32]. For instance, the parameters used for the analysis of the Fe nanowire (Fig. 1(b)) were as follows: saturation magnetization $M_s = 1.2 \times 10^6$ A/m, wire length $l = 12.5$ μm, wire diameter $d = 188$ nm, and cell (mesh) size of $4 \times 4 \times 4$ nm$^3$. Second, we analyzed the field projected onto the axis of the NV centers chosen for the image. Finally, we considered the wide-field imaging conditions and NV measurement parameters to ensure that the simulation reflected real experimental situations. For instance, we had to include the field averaging effect owing to the presence of multiple NV centers within the laser spot of ~1 μm that experienced diverse fields because of the large field gradient near the nanowire[33]. We also had to consider the Airy disk point spread function for the photoluminescence of the NV centers[34,35], as well as the optical beam paths in our TIRF configuration. Combining these effects was necessary to explain the overall reduced field strength compared with the OOMMF result and the hole-like features appearing inside the dipole. Based on a comparison between the experimental and simulation results, we were able to extract data on the saturation magnetization of individual nanowires that are compatible with bulk values, i.e., $M_s = 1.7 \times 10^6$ A/m for Fe and $M_s = 1.4 \times 10^6$ A/m for Co[36], but varies moderately depending on the wire. More information regarding the simulation is provided in the Supplementary Materials.

As presented in Fig. 1(a), we repeated the measurement on multiple Fe nanowires while changing the external magnetic field using an NdFeB permanent magnet. Before we dispersed the nanowires on the diamond surface, we magnetized them along the long axis of the wire with a saturation field of >5000 G. The magnetization direction was confirmed based on the magnetic image recorded at 71 G (not shown here). We then switched the field sign and gradually increased the magnitude from –71 G to –375 G to observe magnetization reversal. The corresponding results are depicted in Figs. 2(b)–(f). Notably, the listed field values corresponded to the field magnitudes along the NV crystal axis, and we used the NV centers with in-plane directions approximately along the long axis of the wire. Figures 2(g)–(k) present close-up images of an example wire marked by the dashed box in Fig. 2(a). The magnetization remained in the same direction at magnetic fields up to –251 G; however, the magnetization was completely reversed above –365 G. At –303 G, we observed incoherent magnetization reversal propagation and the formation of two magnetic domains with approximately equal sizes (Fig. 2(i)). These domains are likely to be transverse[37]. Following the image analysis of Fig. 2(g)–(k), we plotted the normalized magnetization as a function of the external field in Fig. 2(l). The result revealed a dissimilar hysteresis curve compared with the VSM result obtained from the Fe nanowire arrays (inset of Fig. 2(l)). The $H_c$ value of approximately –300 G was several times greater than that of the nanowire arrays (shaded area in Fig. 2(l)), owing to the absence of inter-wire magnetostatic interactions for single nanowires. Notably, magnetization reversal occurs rapidly around coercivity, with only a small field increment. These results suggest the magnetically harder ferromagnetic behavior of a single nanowire compared to that of the ensemble. The hysteresis curves of the other nanowires, illustrated in Fig. 2(a), are depicted in Fig. S6, indicating similar results.

After studying single-component nanowires (Fe and Co), we extended the experiment to BMNs with various material compositions and layer sequences. Notably, identifying the different materials and segments contained in BMNs solely from optical images obtained using conventional microscopes is usually difficult, and high-resolution imaging tools, such as SEM, are required. A key motivation driving our experiment was to identify the materials in each layer and extract their magnetic properties using wide-field diamond microscopy. We first examined BMNs consisting of nonmagnetic (e.g., Au) and ferromagnetic (e.g., Fe) materials. Figure 3 presents the magnetic images of two different types of Fe–Au BMNs: Fe–Au–Fe and Au–Fe–Au. We fabricated both BMNs with a similar length of approximately 20 μm and a diameter of approximately 200 nm but with different sequences. We performed the measurements in a relatively weak magnetic field of ~100 G. While the VSM data of the Au nanowire arrays indicated paramagnetic behavior, we did not observe any magnetic signals at 100 G in the separate imaging experiments conducted on single Au nanowires. This suggests either the nonmagnetic or extremely weak paramagnetic characteristics of single Au nanowires that are not detectable within the sensitivity of our microscope, i.e., ~80 μT/$\sqrt{\text{Hz}}$; this is also clearly displayed in Fig. 3, where dipole-like features appear only at the ends of the Fe segments. For instance, in Fig. 3(a), Fe–Au–Fe appears as two separated short Fe nanowires. In Fig. 3(b), however, the dipole does not appear at the wire tips but only appears at the interfaces between the Fe and Au layers, suggesting the absence of stray fields emanating from Au.

To determine whether we could detect these differences even when the BMNs were made of different ferromagnetic materials, we replaced Au with Co, which is expected to have a smaller magnetization than Fe based on bulk measurements[35]. Similar to the experiments in Fig. 3, we studied two types of Fe–Co BMNs with similar lengths of approximately 20 μm and diameters of approximately 200 nm: Fe–Co–Fe and Co–Fe–Co. As depicted in Figs. 4(a) and (b), the two BMNs were indistinguishable based on their optical images. However, the magnetic images revealed distinct profiles between them. For the Fe–Co–Fe BMN, the field magnitude and dipole size at the interface were noticeably smaller than those at the tips. Moreover, the dipole polarity was reversed. This was not the case for the Co–Fe–Co BMN, where no significant differences were observed between the tips and interfaces in terms of the field strength, dipole size, and polarity. The illustrations in Fig. 4(c) explain this discrepancy qualitatively. When two different ferromagnets are combined in series to form a wire, the relative order between the stronger (e.g., Fe) and weaker ferromagnets (e.g., Co) is essential. At the Fe–Co interface, the stray field from the Fe segment enters the wire, whereas the opposite is true for Co. Because the former is stronger than the latter, the net field at the interface is reduced and points toward the wire body, which is opposite to the field direction at the Fe tip. However, for Co–Fe, the fields at the Co tip and the Co–Fe interface are oriented along the same direction, with comparable strengths. Therefore, unlike optical images, magnetic images enable us to distinguish the relative sequence and the amount of magnetization of different materials contained in the BMNs.

For a more quantitative analysis, we compared the magnetic images with a micromagnetic simulation, as depicted in Fig. 5. Two types of BMNs were used as examples: Fe–Au–Fe and Fe–Co–Fe. Based on independent measurements, we determined most simulation parameters in advance, including the experimental parameters of the NV center and wide-field microscopy. For instance, the experimental parameters employed for the simulation in Fig. 5 were as follows: average of one NV center per $20 \times 20$ nm$^2$, an NV depth of 15 nm, a diffraction-limited laser spot of 1 µm, an ODMR linewidth of 6 MHz, and an ODMR contrast at a zero field of 1%. For the structural parameters of the wire, we obtained the total length from the optical image and the length of each segment from the magnetic image, that is, from the distance between the dipoles. For instance, the parameters employed for Fig. 5 were as follows: (a) a total wire length of 15 µm and Fe/Au layer length of 2.4 µm/10.2 µm and (b) a total wire length of 18 µm and Fe/Co layer length of 6.5 µm/5 µm. With these predetermined parameters, we ran simulations with various values of the saturation magnetization $M_s$ and wire diameter $d$ until the simulated image matched the measured image in terms of the field magnitude and dipole size. The best simulation parameters for Fig. 5(a) were $M_s$ (Fe) = $1.2 \times 10^6$ A/m and $d$ = 172 nm, indicating ~5% and ~10% discrepancies between the measurement and simulation, respectively, in terms of the maximum field strength and dipole size. From Fig. 5(b), however, we obtained the following: $M_s$ (Fe) = $1.2 \times 10^6$ A/m, $M_s$ (Co) = $1.0 \times 10^6$ A/m, and $d$ = 180 nm, indicating ~6% (field strength) / ~8% (dipole size) deviations at the tips and ~33% (field strength) / ~30% (dipole size) deviations at the interfaces. The obtained $M_s$ values were smaller than those of the bulk metals. Based on the separate measurements (10 nanowires for Fe and Co each), we identified a similar reduction from ~60% of the nanowires. Previous studies have also reported reduced $M_s$ values for single nanowires[2,23,38].

## Conclusion

Herein, we demonstrated a novel imaging method for characterizing individual magnetic nanowires and BMNs under ambient room temperature conditions. We analyzed the magnetic properties of these nanowires, such as the saturation magnetization and coercivity, using wide-field diamond microscopy at the single-nanowire level. We obtained magnetic hysteresis curves from the evolution of magnetization under various external fields. The adopted microscope enabled us to simultaneously study multiple nanowires within the field of view. Moreover, we identified different material components, their relative locations, and sequences within a single BMN. The method introduced in this study is relatively simple and easy to operate and, thus, provides a versatile tool for various applications based on magnetic nanostructures. We expect that further improvements in the current experimental setup, such as those facilitating the application of a larger external field with finer steps or allowing arbitrary changes in the field direction, will more precisely elucidate the magnetic properties of individual nanowires and BMNs such as the mechanism of magnetization reversal.

## Acknowledgments

We acknowledge financial support from the National Research Foundation of Korea (2018M3C7A1024602, 2019R1A2C3006587, and 2021R1F1A1049355). The research related to the heat treatment of single-crystal diamonds was supported by the Institute for Basic Science (IBS-R019-D1).

## Contributions

JY, JHM, and JJ conceived the project. JY, JJ, KK, and DL constructed the wide-field quantum microcopy setup, conducted the imaging experiments, and analyzed the data. SHL, HSK, and C.-H.L. annealed and chemically cleaned the diamond plates. JHM and EO fabricated the magnetic nanowires and conducted M–H loop measurements on the nanowire arrays. YJK, YSJ, and YKK analyzed the magnetic properties. KH, DL, YKK, and DL directed and supervised the project. All authors contributed to the discussion and preparation of the manuscript.

**Methods**

**Synthesis of magnetic nanowires and BMNs**

We fabricated Fe and Co nanowires and Fe–Au and Fe–Co BMNs using porous AAO templates containing 200 nm diameter pores via electrodeposition. We deposited 300 nm thick Ag films using an e-beam evaporator on one side of the AAO templates to act as working electrodes. A Pt plate was used as the counter electrode. Each solution was prepared independently and changed after the growth of one segment to prevent the formation of an alloy and maintain the characteristics of the pure metal. We dissolved iron sulfate heptahydrate (FeSO$_4$·7H$_2$O, 80 mM), cobalt (II) sulfate heptahydrate (CoSO$_4$·7H$_2$O, 80 mM), and potassium dicyanoaurate (I) (KAu(CN)$_2$, 8 mM) in deionized (DI) water and added boric acid (H$_3$BO$_3$, 400 mM) to serve as a buffer solution for all the solutions. A source meter (Keithley 2612 B) was used to apply current densities of 2.5 mA/cm$^2$ for Fe and Co and 0.25 mA/cm$^2$ for Au. We used different lengths for each segment ranging from 3 μm to 20 μm. We targeted to fabricate Fe–Au–Fe (5 μm–9 μm–5 μm), Au–Fe–Au (3 μm–14 μm–3 μm), Fe–Co–Fe (4 μm–12 μm–4 μm), Co–Fe–Co (6 μm–8 μm–6 μm), Fe (15 μm), Co (18 μm), and Au (20 μm) nanowires by adjusting the growth rate for each segment: 4.0 nm/s for Fe, 3.4 nm/s for Co, and 0.7 nm/s for Au. We magnetized the nanowires embedded in the AAO templates using a permanent magnet. Subsequently, we removed the Ag layer using an etchant, followed by the removal of the AAO template at 45 ℃ for 15 min using a sodium hydroxide (NaOH, 3 M) solution. We washed the template residue and NaOH several times using DI water and dispersed the nanowires in ethanol.

**Analysis of the bulk nanowires**

We examined the morphology of the nanowires using SEM (VEGA 3, Tescan). First, we distinguished the length of each layer based on the difference in the contrast of SEM images (Figs. S7 and S8). However, Co and Fe, which have similar atomic numbers, did not show a clear contrast at the boundary (Fig. S8). Next, we double checked the wire dimensions using optical microscopy. Following this, we analyzed the sample microstructure using X-ray diffraction (Aeris, Malvern Panalytical). Finally, we analyzed the magnetic properties of the nanowire array embedded in the AAO templates using VSM (Microsense EV9). We performed all measurements in the presence of a magnetic field at room temperature and normalized the magnetic moment to the maximum value within the measurement range.

**Preparation of diamond and NV centers**

We used commercially available electronic-grade diamond plates from Element Six ($2 \times 2 \times 0.5$ mm$^3$) and thinned them to $100$ μm for the TIRF experiment. The diamond plates were implanted with $^{15}$N$^+$ ions at 10 keV at a density of $10^{14}$ cm$^{-2}$ and annealed at 1200 °C under high vacuum for 2 h. The average depth of the NV centers was estimated to be ~15 $\pm$ 5 nm[39]. Next, we performed oxygen annealing at 465 °C for 4 h under an oxygen flow of 200 sccm to stabilize the charge state of NV[39,40,41].

**Wide-field diamond microscopy setup and ODMR**

We used a high-power 532 nm laser at 500 mW to excite the ensemble of NV centers. To avoid unwanted heating of the nanowires, we ran our experiment in the TIRF configuration with a commercial TIRF objective lens (CFI Apochromat TIRF 60XC Oil). We used the objective lens for both excitation and collection of the fluorescence signals of the NVs, which was further detected by the scientific CMOS camera (pco, panda 4.2).

**Data availability**

The data that support the findings of this study are available from the corresponding authors upon reasonable request.

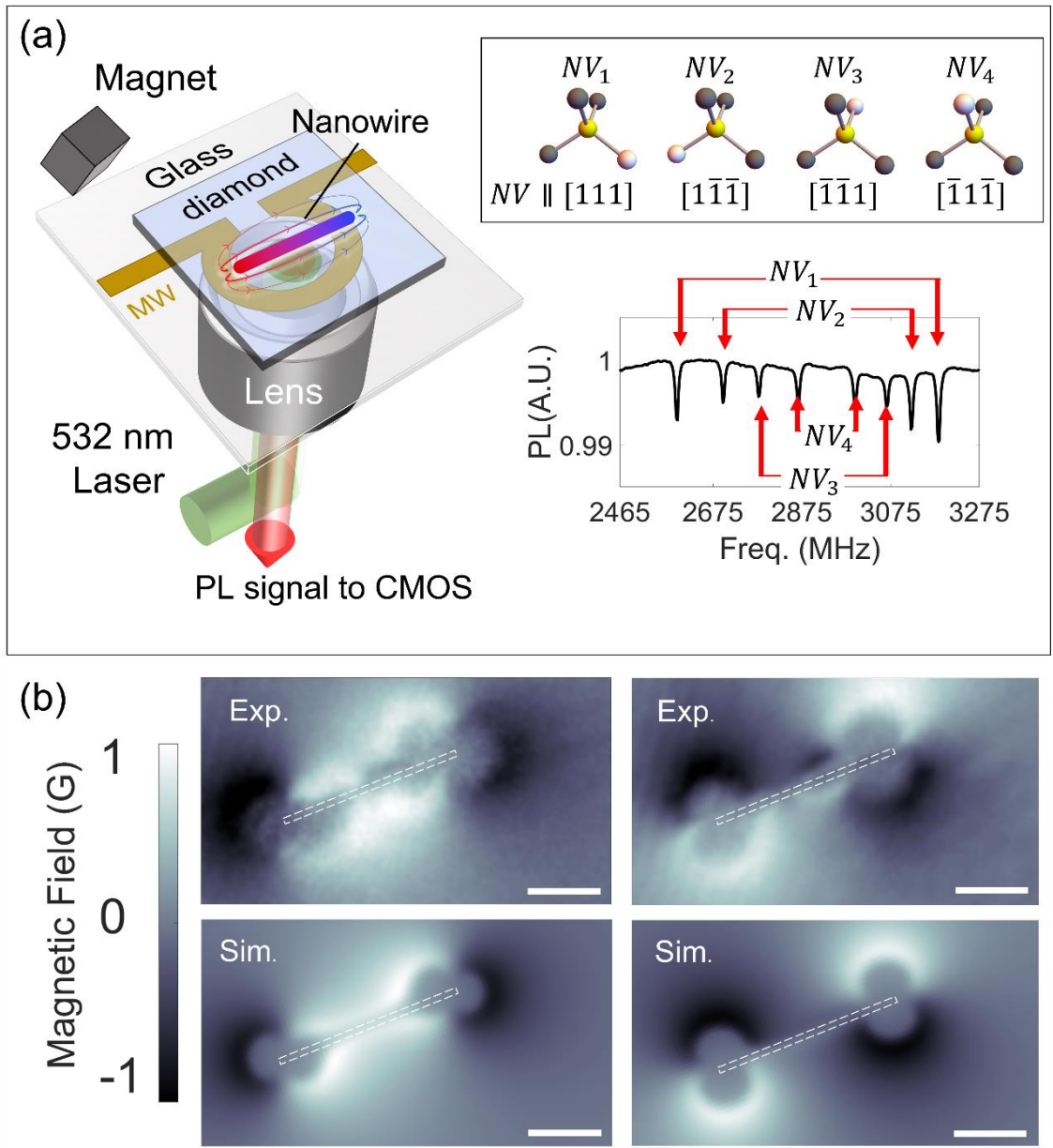

**Fig. 1. Working principle of magnetic imaging with diamond NV centers.** (a) Schematic of the wide-field diamond microscope. A thin diamond plate ($2 \times 2 \times 0.1$ mm$^3$) hosts an ensemble of NV centers at an average depth of 15 nm from the surface, where magnetic nanowires or BMNs are located. The NV centers are excited using a 532 nm laser, and the PL signals are recorded and imaged using a CMOS camera. An omega-shaped microwave strip line is used to drive the transition between spin states that is optically detectable, thus enabling the ODMR measurement. Owing to the existence of four possible crystal axes of the NV centers, we can analyze the magnetic field along four different directions. (b) Examples of magnetic images of an Fe nanowire obtained along two different axes of the NV centers, i.e., NV$_1$ ∥ [111] and NV$_3$ ∥ [$\bar{1}\bar{1}1$]. We compared the experimental results with those of a micromagnetic simulation. The dipole-like features are the signatures of the stray fields from the wire that are the largest at the tips. The dipoles are aligned along the in-plane direction of the NV axis. The dashed rectangles indicate the nanowire. The scale bar = 5 μm.

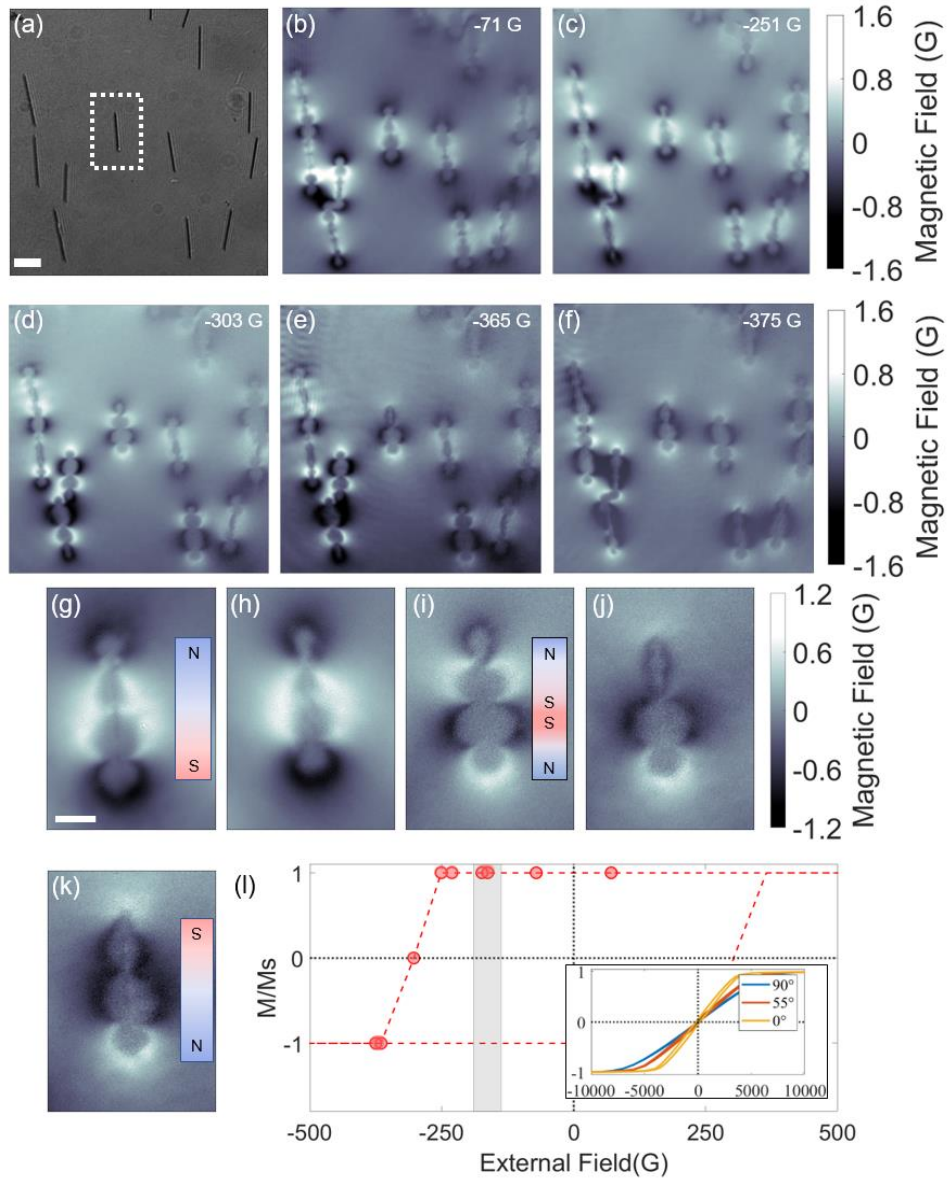

**Fig. 2. Magnetic hysteresis measurement of an individual Fe nanowire.** Optical image (a) and magnetic images (b)–(f) of single Fe nanowires within the field of view (scale bar = 10 μm). We obtained the magnetic images at various external magnetic fields by moving a permanent magnet. Before the experiment, we magnetized the nanowires up to the saturation magnetization at >5000 G. Then, we flipped the field direction and gradually increased the magnitude to observe the magnetic reversal of each wire. (g)–(k) Magnetic images of the Fe nanowire marked in the dashed box in (a). The scale bar = 5 μm. The evolution of the field profiles clearly depicts the magnetization reversal between (g) and (k). The image in (i) displays the splitting of the magnetic domain into two, suggesting net zero magnetization. We obtained the magnetic hysteresis curve depicted in (l) by extracting the normalized magnetization data as a function of the external field along the NV axis, whose in-plane direction was parallel to the long axis of the wire. The inset depicts the VSM data obtained from Fe nanowire arrays at three field angles. The shaded area in (l) indicates the range of coercivity obtained from the VSM measurement.

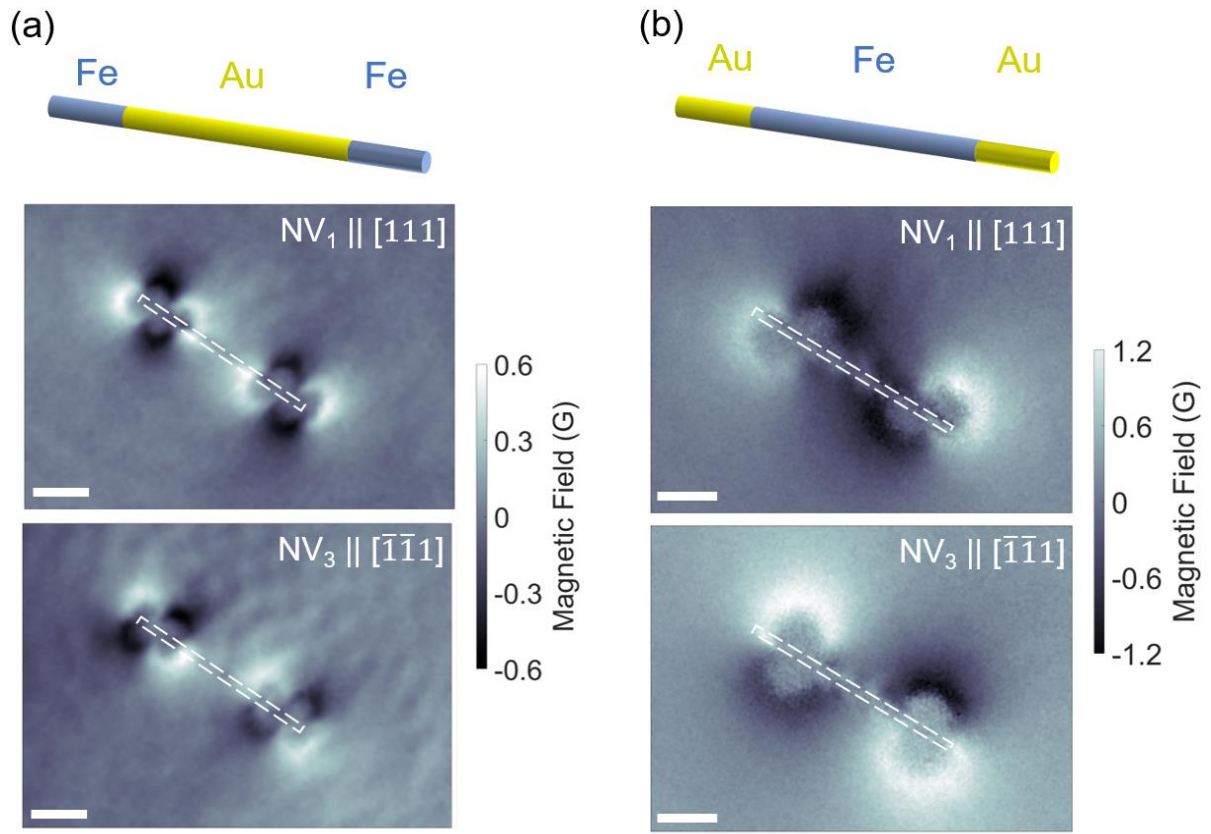

**Fig. 3. Magnetic images of BMNs consisting of ferromagnetic (Fe) and nonmagnetic (Au) segments.** Magnetic images of the (a) Fe–Au–Fe and (b) Au–Fe–Au BMNs obtained using $NV_1 \parallel [111]$ and $NV_3 \parallel [\bar{1}\bar{1}1]$. Dipole features appear only at the ends of the Fe segments, suggesting the nonmagnetic or weak paramagnetic nature of Au. The scale bar = 5 μm.

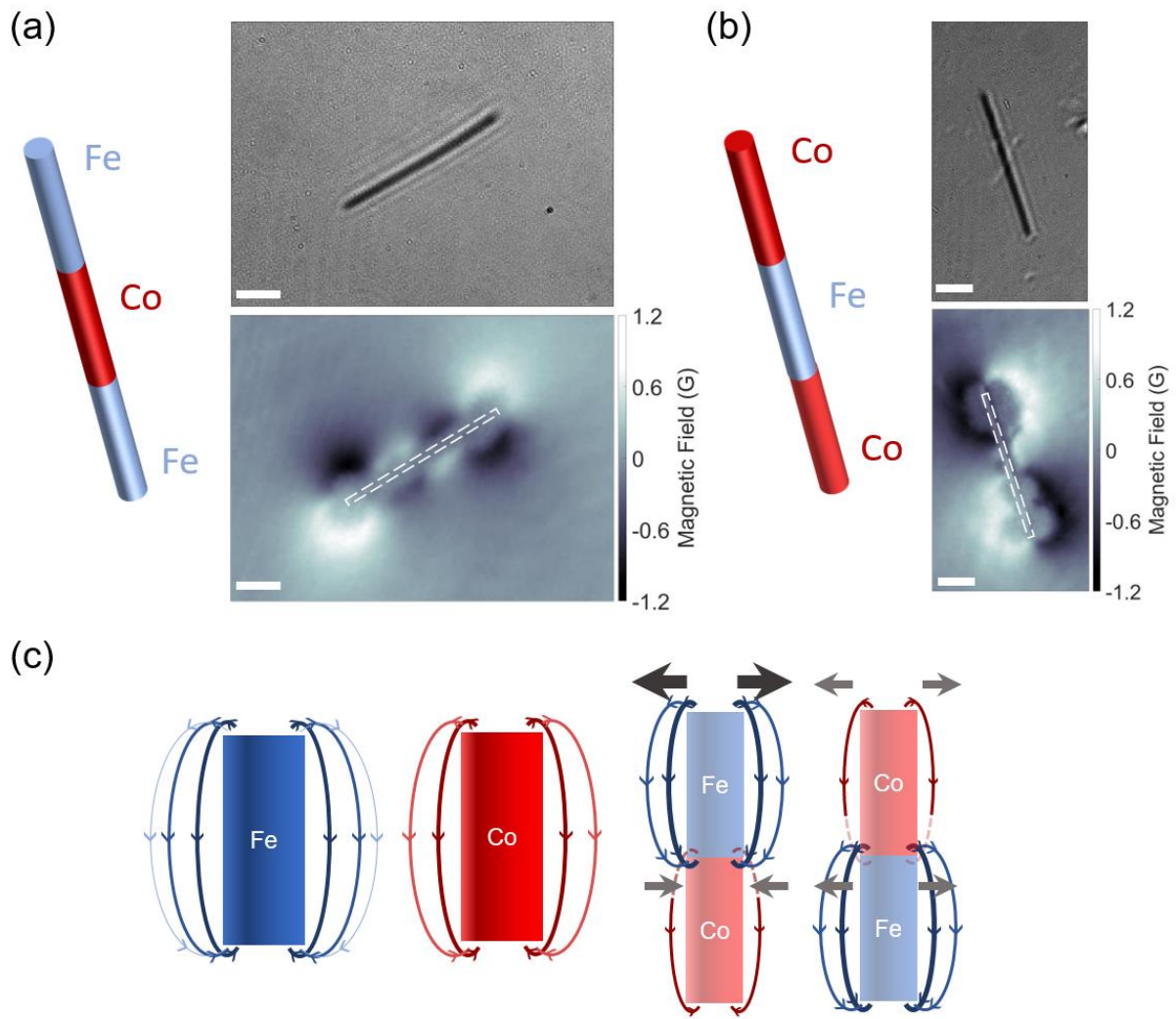

**Fig. 4. Magnetic images of the BMNs consisting of two different ferromagnetic segments, Fe and Co.** Optical and magnetic images of the (a) Fe–Co–Fe and (b) Co–Fe–Co BMNs. The scale bar = 5 µm. In contrast to the optical images, the magnetic images display distinct signatures. The stray field profiles at the tips and the interfaces for Fe–Co–Fe are dissimilar in terms of the field strength, polarity, and dipole size. However, the Co–Fe–Co profiles are almost the same. Schematics in (c) explain this difference. When the Fe and Co segments are connected in series to form a nanowire, the net magnetic field at the interface varies depending on the order of the segments. For instance, the Fe–Co configuration explains the reduced magnitude and reversed sign of the field at the interfaces in (a). As shown in (b), this is not the case for the Co–Fe configuration, where the field magnitude and the sign can be almost identical.

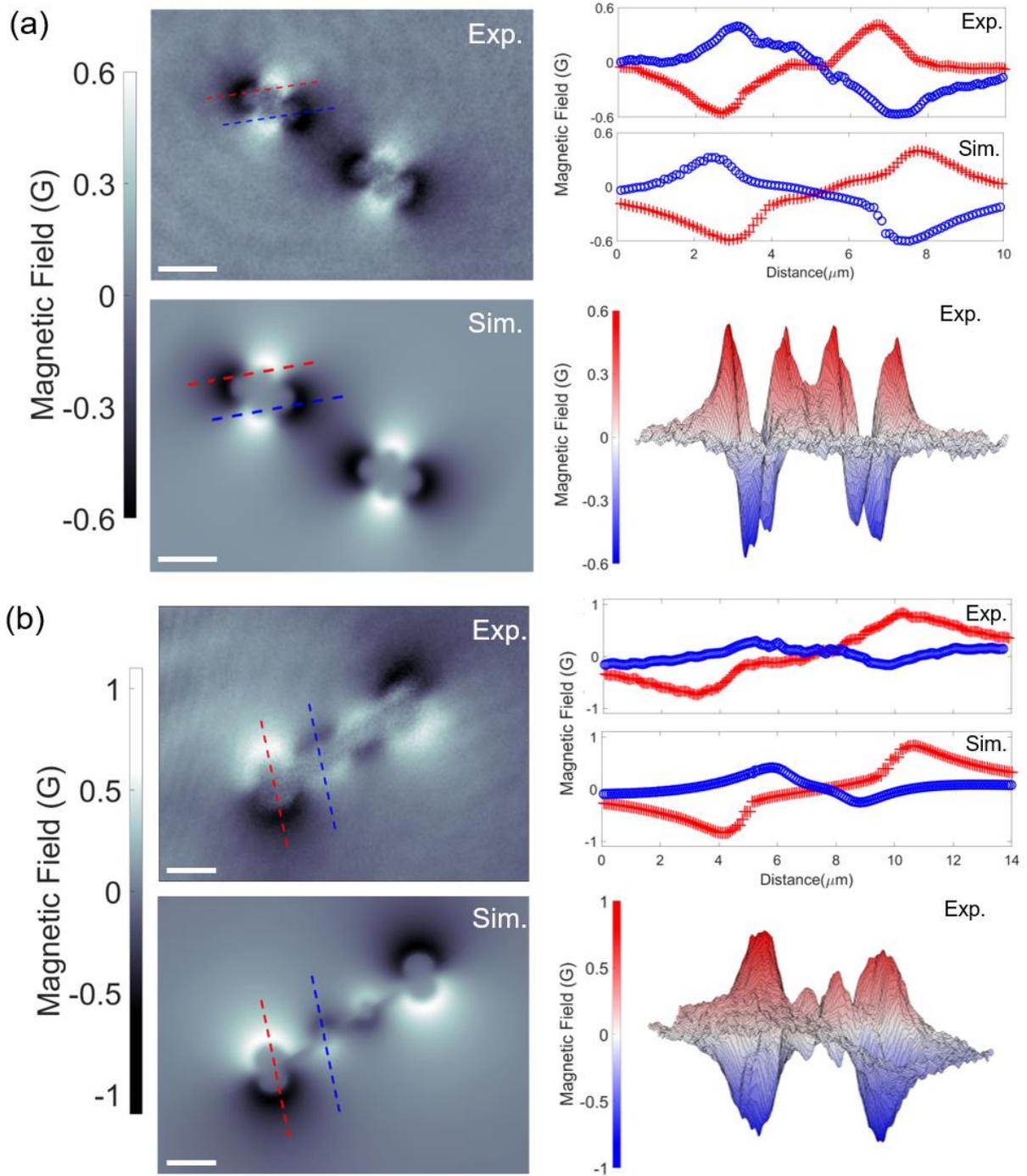

**Fig. 5. Magnetic analysis from the comparison between experimental and simulation data.**
Experiment and simulation images of (a) Fe–Au–Fe and (b) Fe–Co–Fe (b) previously shown
in Fig. 3(a) and Fig. 4(a). The scale bar = 5 μm. We tuned the simulation parameters, i.e., the
saturation magnetization of each material and wire diameter, until the simulated image matched
with the measured image in terms of the field magnitude and the dipole size. The line-cut data
compare the magnetic profiles along the dashed lines marked in the images. The three-
dimensional profiles are also plotted to visualize the local maxima of the magnetic field at the
tips and interfaces of the wires.

# Supplementary Materials

## Exploring the magnetic properties of individual barcode nanowires using wide-field diamond microscopy

Jungbae Yoon[1,+], Jun Hwan Moon[2,+], Jugyeong Jeong[1], Yu Jin Kim[3], Kihwan Kim[1], Hee Seong Kang[4], Yoo Sang Jeon[5], Eunsoo Oh[2], Sun Hwa Lee[6], Kihoon Han[7,8], Dongmin Lee[7,9], Chul-Ho Lee[4], Young Keun Kim[2,3*] and Donghun Lee[1,*]

[1]Department of Physics, Korea University, Seoul, Republic of Korea
[2]Department of Materials Science and Engineering, Korea University, Seoul, Republic of Korea
[3]Institute for High Technology Materials and Devices, Korea University, Seoul, Republic of Korea
[4]KU-KIST Graduate School of Converging Science and Technology, Korea University, Seoul, Republic of Korea
[5]Center for Hydrogen Fuel Cell Research, Korea Institute of Science and Technology, Seoul, Republic of Korea
[6]Center for Multidimensional Carbon Materials (CMCM), Institute for Basic Science (IBS), Ulsan 44919, Republic of Korea
[7]BK21 Graduate Program, Department of Biomedical Sciences, Korea University College of Medicine, Seoul, Republic of Korea
[8]Department of Neuroscience, Korea University College of Medicine, Seoul, Republic of Korea
[9]Department of Anatomy, Korea University College of Medicine, Seoul, Republic of Korea
[+] These authors contributed equally to this work.

*Corresponding authors:
Email addresses: ykim97@korea.ac.kr (Y. K. Kim) and donghun@korea.ac.kr (D. Lee)

### Note 1. Experimental setup

The wide-field diamond microscope is illustrated in Fig. S1. An excitation laser with a wavelength of 532 nm was focused onto nitrogen-vacancy (NV) ensembles through a total internal reflection fluorescence (TIRF) objective lens with a field of view of ~100 µm. The NV centers were located approximately 15 nm below the surface of a diamond plate ($2 \times 2 \times 0.1$ mm$^3$). For wide-field optical images, the fluorescence signals of the NV centers were recorded using a scientific complementary metal–oxide–semiconductor camera placed after a dichroic mirror. For the spin transitions of the NV centers, a microwave with a frequency of

~2.9 GHz was generated by the omega shape of a gold strip line. To disperse the magnetic nanowires or barcode magnetic nanowires (BMNs) over the diamond surface, we placed a droplet of a methanol solution containing the nanowires onto the diamond surface and waited until it dried.

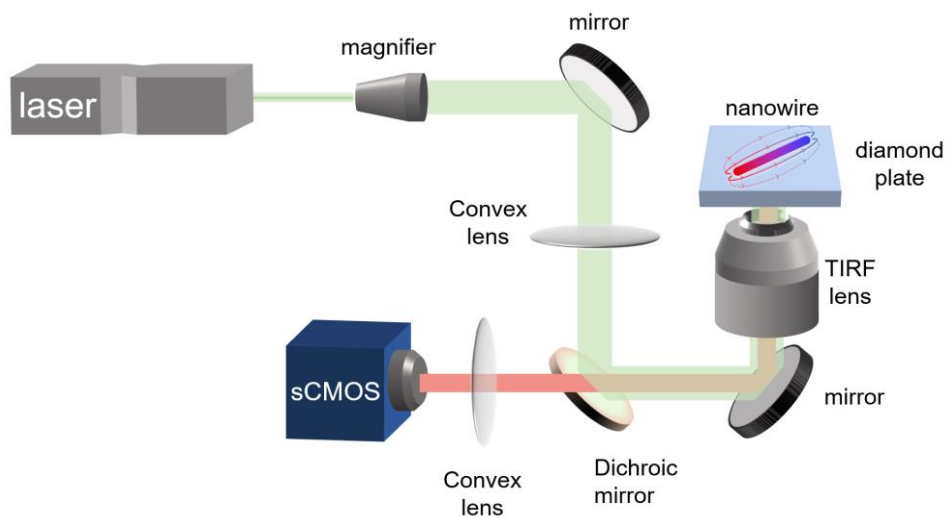

**Fig. S1**. Schematic of the wide-field diamond microscope setup.

## Note 2. Optically detected magnetic resonance (ODMR) measurement

Notably, a negatively charged NV center in diamond exhibits S = 1 triplet spin states at the ground energy level. At room temperature, the degenerated $m_s = \pm1$ states lie approximately 2.9 GHz apart from the $m_s = 0$ state owing to the zero-field splitting. Upon illumination with the 532 nm laser, the NV center at the $^3A_2$ ground orbital states can be optically excited to the $^3E$ excited orbital states, as depicted in Fig. S2(a). In general, the NV center immediately (~10–15 ns) returns to the ground state by emitting broadband photons with wavelengths of 637 nm to 800 nm. However, there exists a finite probability (~30 %) for the NV center to follow a different relaxation route involving $^1A_1$ and $^1E$ singlet states only when the spin states are initially either $m_s = +1$ or $m_s = -1$. This channel is accompanied by relatively slow (~400–500 ns) and nonradiative relaxation, as well as spin-flip transitions to $m_s = 0$. Therefore, the unique optical transition provides spin initialization to $m_s = 0$ upon long-enough excitation with the 532 nm laser (e.g., ~1 μs); moreover, the spin-dependent photoluminescence (PL) signal, that is, the signal corresponding to $m_s = 0$, is greater than that of $m_s = \pm1$. This is the underlying mechanism of the ODMR measurement. Figure S2(b) illustrates a schematic of the ODMR measurement, depicting a reduction in the PL signal at the frequency of spin transitions corresponding to either $m_s = 0 \leftrightarrow m_s = -1$ or $m_s = 0 \leftrightarrow m_s = +1$. A non-zero magnetic field results in the Zeeman splitting of the degenerated $m_s = \pm 1$ states by $2\gamma_e B_{\parallel NV}$, where $\gamma_e$ denotes the gyromagnetic ratio of the NV spin, and $B_{\parallel NV}$ represents the magnetic field parallel to the crystal axis of the NV center. The magnetic field image analyzed in this study was obtained by recording the amount of the Zeeman shift of the NV centers in every pixel of the image.

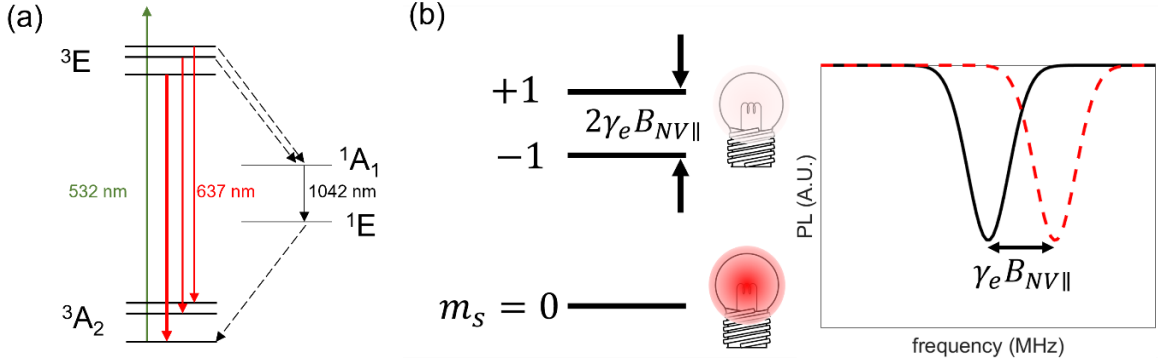

**Fig. S2**. Mechanism of the ODMR measurements of NVs. Schematics of (a) the optical transition of the NVs and (b) the ground energy levels of the NVs, illustrating the Zeeman splitting of the $m_s = \pm 1$ states owing to the external magnetic field along the NV axis, $B_{\parallel NV}$. The ODMR measurement presents a reduction in the PL signal at the frequency of the spin transitions, for example, at $m_s = 0 \leftrightarrow m_s = -1$, and we obtain $B_{\parallel NV}$ from the amount of the Zeeman shift.

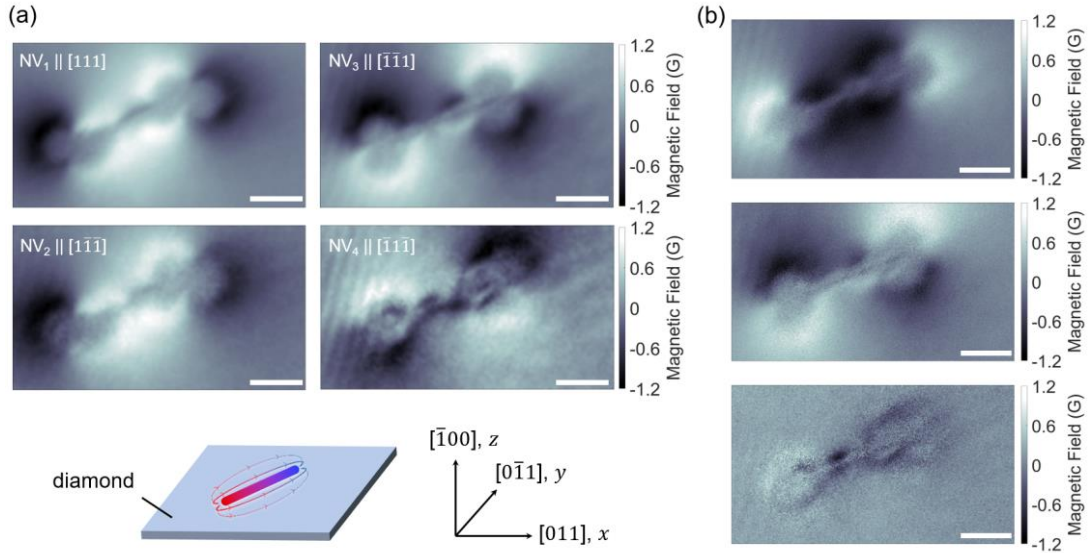

**Fig. S3**. Example of vector magnetometry. (a) Magnetic images of an Fe nanowire obtained using NV centers with four different crystal axes. (b) Converted magnetic images, with field components along the *x*, *y*, and *z* directions based on the coordinate system shown in (a). The scale bar = 5 μm.

**Note 3. Magnetic images of a single Fe nanowire**

As discussed in the main text, we can realize vector magnetometry by analyzing the magnetic images obtained along different axes of the NV centers. An example of vector magnetometry is presented in Fig. S3. Figure S3(a) depicts magnetic images captured along four different field directions parallel to the NV axes, that is, $NV_1 \parallel [111]$, $NV_2 \parallel [1\bar{1}\bar{1}]$, $NV_3 \parallel [\bar{1}\bar{1}1]$, and $NV_4 \parallel [\bar{1}1\bar{1}]$. The first and third images are used in Fig. 1(b) of the main text. Based on a combination of the data displayed in Fig. S1(a), we computed the field components along the $[011]$, $[0\bar{1}1]$, and $[\bar{1}00]$ crystal axes corresponding to the $x$, $y$, and $z$ directions of the coordinate system. The converted magnetic images of $B_x$, $B_y$, and $B_z$ are presented in Fig. S3(b).

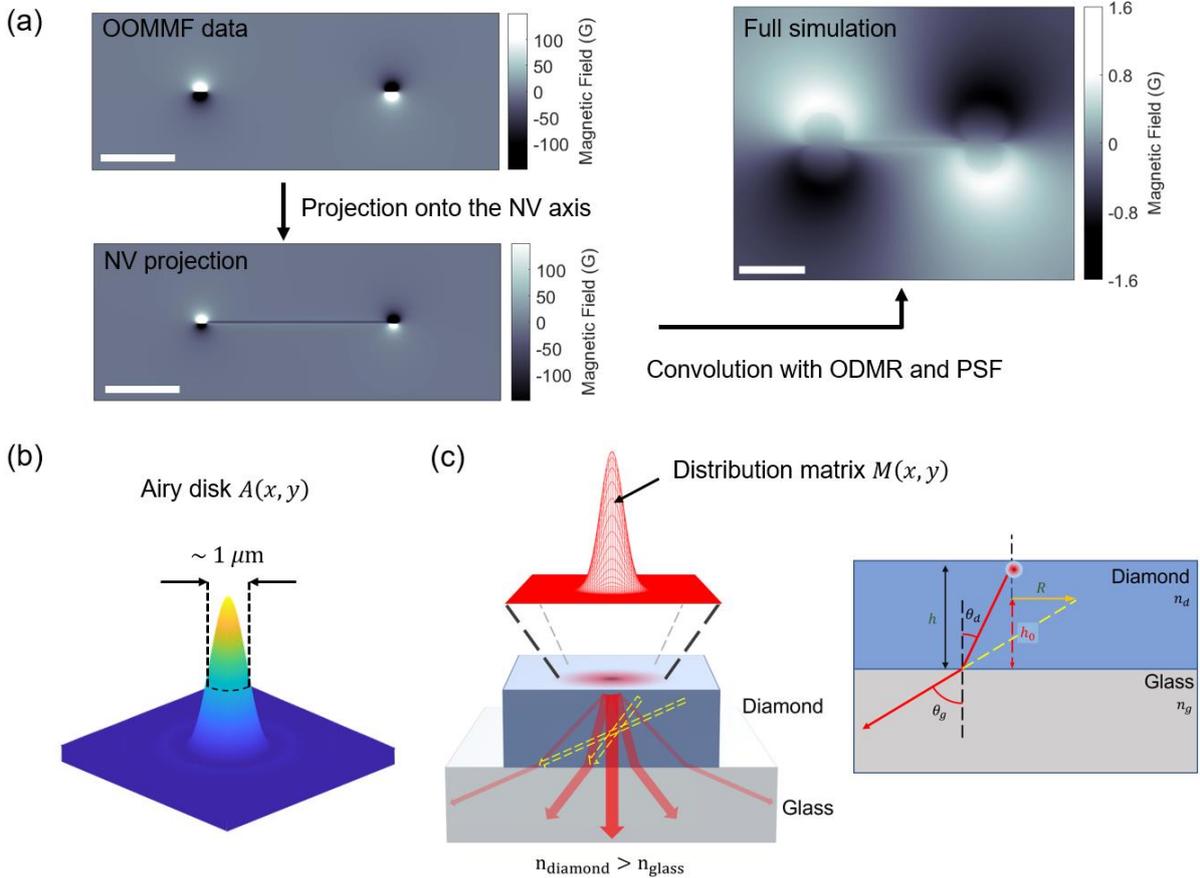

**Fig. S4**. Simulation procedure. (a) Three main steps involved in the simulation: Object oriented micromagnetic framework (OOMMF) simulation, projection of the OOMMF field onto the NV axis, and convolution with the ODMR and point spread function (PSF). The scale bar = 5 μm. (b) Airy disk PSF, $A(x, y)$, used to accommodate the single emitter nature of the NV center. (c) Distribution matrix, $M(x, y)$, numerically obtained by considering the refraction of the emitted photons at the interface between diamond and glass.

**Note 4. Simulation method**

The simulation procedure is illustrated in Fig. S4(a). First, using the OOMMF, we numerically simulated the magnitude and direction of the magnetic field around a nanowire or BMN. We used a mesh size of $4 \times 4 \times 4$ nm$^3$ and the following magnetic values for the nanowires: anisotropy constant $K_1 = 4.7 \times 10^4$ J/m$^3$ and exchange energy $A = 25$ pJ/m for Fe (BCC crystal structure), as well as $K_1 = 4.7 \times 10^4$ J/m$^3$ and $A = 25$ pJ/m for Co (HCP crystal structure). Note that, in this study, we used the saturation magnetization $M_s$ as a variable and manually changed the value until we noticed good agreement between the simulation and measurement results. Second, we extracted the field component whose direction matched the axis of the NV centers used for the image. This was achieved by projecting the OOMMF data onto the axis. Finally, we obtained the complete simulation result by considering the ODMR measurements and wide-field imaging conditions.

As illustrated in Fig. S2(b), the amount of shift in the ODMR was used as a measure of the magnetic field in each pixel of the image. In our setup, the resonance peak consisted of contributions from multiple NV centers within the pixel, which may experience different magnetic fields. This is particularly true for pixels near the magnetic nanowire, where the field gradient is the largest. Therefore, we included the field gradient effect and constructed an averaged ODMR field. We first obtained the ODMR data for a single NV center using the Gaussian resonance given in Eq. 1.

$$I(x, y, f) = C \left[ 1 - \exp\left[ -\frac{(f - f_\pm)^2}{2\sigma^2} \right] \right] \tag{1}$$

Here, the normalized PL intensity $I$ was plotted as a function of the NV position in the image $(x, y)$ and the microwave frequency $f$. Based on the NV density, we assumed that, on average, a single NV center is located every $20 \times 20$ nm$^2$. Here, $C$ and $\sigma$ denote the ODMR contrast and linewidth, which were determined to be 1% and 6 MHz, respectively, from separate measurement data. $f_\pm$ represents the resonance frequency between $m_s = 0$ and $m_s = \pm 1$, which is calculated using Eq. 2.

$$f_\pm(x, y) = D \left[ 1 \pm \left( \frac{\gamma_e}{h} \frac{B(x,y)}{D} \right) \cos(\theta_B(x, y)) + \frac{3}{2} \left( \frac{\gamma_e}{h} \frac{B(x,y)}{D} \right)^2 \sin^2(\theta_B(x, y)) \pm$$

$$\left( \frac{\gamma_e}{h} \frac{B(x,y)}{D} \right)^3 \left( \frac{1}{8} \sin^3(\theta_B(x, y)) \tan(\theta_B(x, y)) - \frac{1}{2} \sin^2(\theta_B(x, y)) \cos(\theta_B(x, y)) \right) \right] \tag{2}$$

Here, $D = 2.87$ GHz represents the zero-field splitting, $h$ is the Planck constant, $B$ is the magnetic field, and $\theta_B$ is the angle between the magnetic field and the NV axis. Notably, $B$ includes the magnetic field contributions from both the permanent magnet and nanowire, and $\theta_B$ is calculated based on the angle between the total field and the NV axis. Both $B$ and $\theta_B$ are functions of the NV location $(x, y)$.

After obtaining the ODMR data for each NV center, we combined these resonances based on the optical imaging conditions employed in the experiment. In every image pixel, we recorded the fluorescence signal from all NV centers within a diffraction-limited spot size of ~1 μm. Because each NV center acts as a point light source, we had to consider the PSF to accommodate the light spreading effect. In this study, we used a PSF based on the Airy disk function with a full width at half maximum of 1 μm, as indicated in Fig. S4(b) and Eq. 3, where $J_1$ denotes the Bessel function of the first kind.

$$A(x,y) = \left[ 2 \frac{J_1\left(\sqrt{x^2+y^2}\right)}{\sqrt{x^2+y^2}} \right]^2 \tag{3}$$

In addition to the Airy disk PSF, we had to consider a TIRF configuration that could deform the collection beam path after passing through materials with different refractive indices, as indicated in Fig. S4(c). In our optical measurement, the fluorescence signal of the NV centers was collected using an oil-immersion TIRF objective lens placed in front of a 100 μm thick cover glass. Owing to refraction at the interface between the diamond and glass, the optical path was modified as if the emission originated from the laterally shifted position. Using Snell's law and the refractive indices of diamond $n_d = 2.42$ and glass $n_g = 1.52$, we determined the relation between the incident angle of the emitted photons $\theta_d$ and the shifted distance $R$, as presented in Eq. 4. The thickness of the diamond layer was $h = 100$ μm.

$$R = h \frac{n_g}{n_d} \tan\left[ \sin^{-1}\left( \frac{n_d}{n_g} \sin\theta_d \right) \right] - h \tan\theta_d \tag{4}$$

We assumed that the photons originating from a single emitter spread uniformly and radially across the diamond sample. We then numerically computed the photon distribution matrix, $M(x,y)$, by counting the relative number of emitted photons at a distance $R(\theta_d)$ while changing the incident angle from $\theta_d = 0^o$ to $\theta_d = 79^o$. The range of $\theta_d$ was determined based on the numerical aperture (NA) of the objective lens (NA = 1.49).

Consequently, we used the modified PSF $P(x,y) = A(x,y) \otimes M(x,y)$, and from the convolution of the modified PSF and the ODMR spectrum for each NV center, we obtained the full averaged ODMR using Eq. 5.

$$I_{\text{full}}(x, y, f) = \iint I(x', y', f) P(x - x', y - y') \mathrm{d}x' \mathrm{d}y' \tag{5}$$

Finally, we extracted the amount of Zeeman shifts from the Gaussian fit results of the simulated $I_{\text{full}}(x, y, f)$ data and obtained the final simulation image, depicted in Fig. S4(a). We found interesting features in the final image that clearly differed from the projected OOMMF image. For instance, the large magnetic field (e.g., >100 G) at the tip of the nanowire was considerably reduced in the final simulation. This can be attributed to two main reasons: first, the detection frequency window used for our ODMR measurement was $\pm$ 15 MHz around the center resonance frequency, which was measured before the introduction of nanowires. Therefore, the ODMRs located outside the window owing to the large Zeeman shift were not accounted for in Eq. 5. Second, extremely close to the tips, the overall ODMR data can include data corresponding to the opposite sign of the magnetic field, resulting in a reduced Zeeman shift after averaging. This is clearly evident in the line-cut data displayed in Fig. 5, where the magnetic field from the tip gradually increased, reached a maximum, and decreased again as it moved away from the tip. This also explains the hole-like features observed in the magnetic images. An example of the simulation results is presented in Fig. S5, which depicts the magnetic field, ODMR contrast, and ODMR linewidth images. We noticed good agreement between the simulation and experimental results.

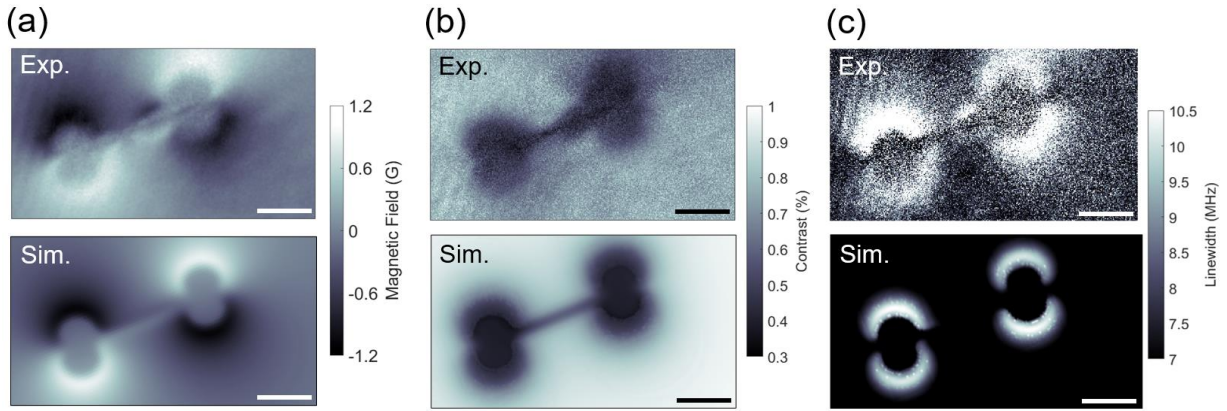

**Fig. S5**. Comparison between the experimental and simulation results. (a) Magnetic field image, (b) ODMR contrast image, (c) ODMR linewidth image. The scale bar = 5 μm.

**Note 5. Magnetic hysteresis plots for the Fe nanowires**

Figure S6 presents the magnetic hysteresis results for the Fe nanowires depicted in Fig. 2 of the main text. Based on the hysteresis data of the 10 nanowires, we discovered that the coercivity values, $H_c$, of individual Fe nanowires ranged from 250–370 G, which are clearly greater than those of the bulk nanowires, that is, nanowire arrays.

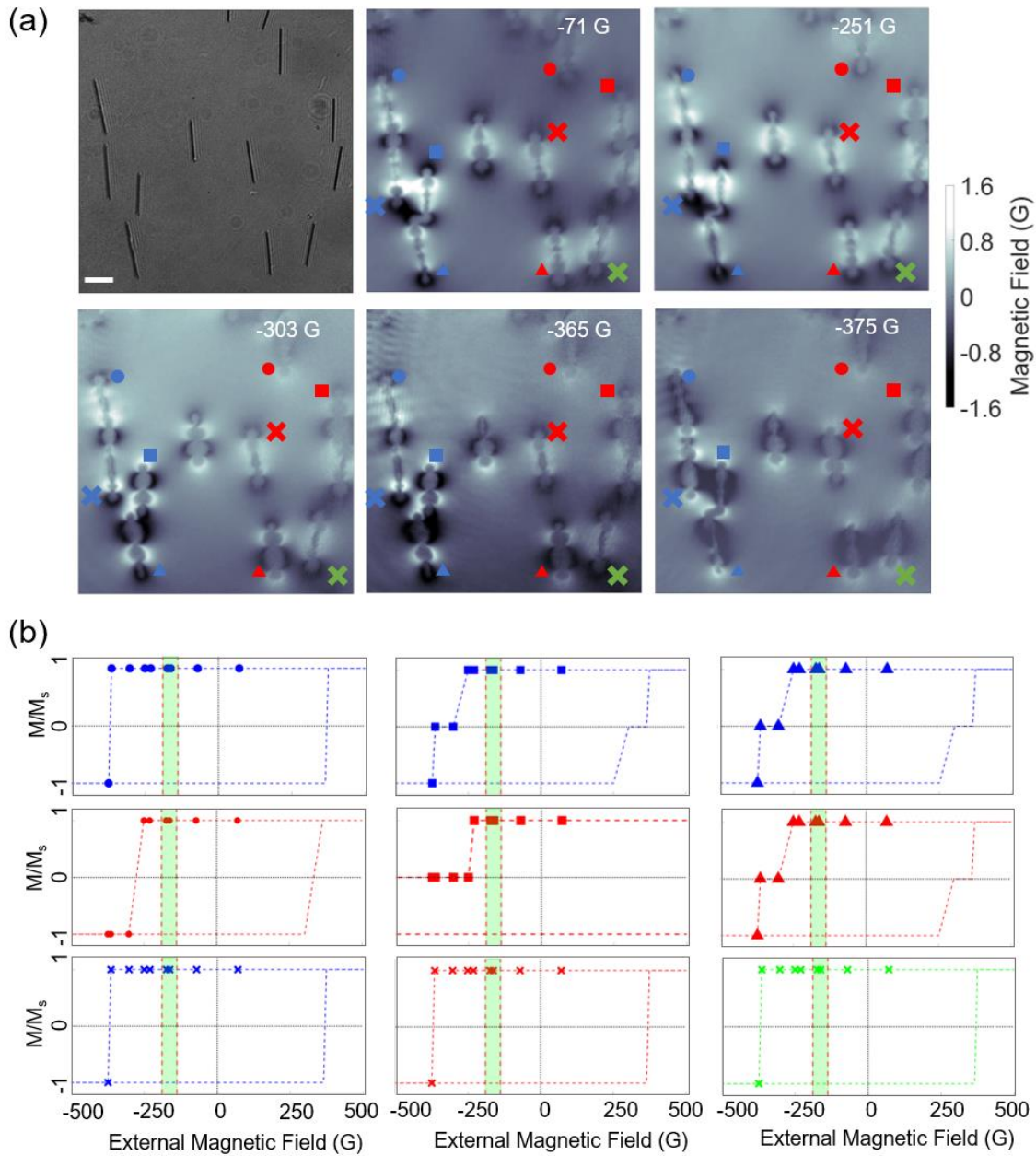

**Fig. S6**. Magnetic hysteresis results for Fe nanowires. (a) Magnetic images of Fe nanowires for various external magnetic fields. The scale bar = 5 µm. (b) Magnetic hysteresis plots for the nine nanowires marked in (a). The shaded area indicates the coercivity values for Fe nanowire arrays.

**Note 6. Scanning electron microscopy (SEM) images and vibrating sample magnetometry (VSM) data**

Figure S7 presents the SEM images and VSM data corresponding to the Co, Fe, and Au nanowire arrays. Figure S8 displays the SEM images of the Au–Fe–Au, Fe–Au–Fe, Co–Fe–Co, and Fe–Co–Fe BMN arrays.

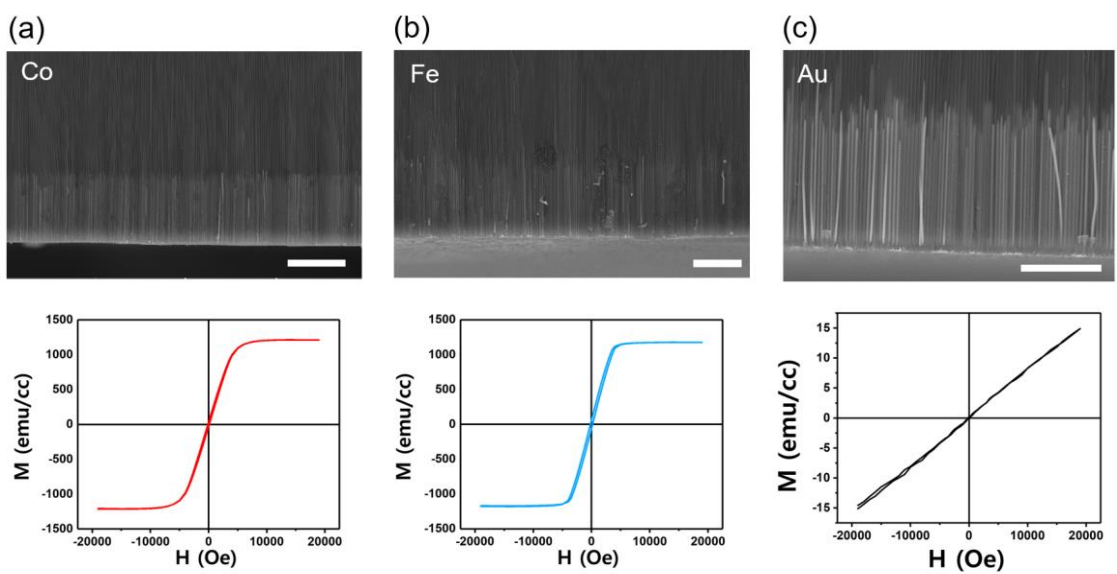

**Fig. S7**. SEM images and VSM data of the (a) Co, (b) Fe, and (c) Au nanowire arrays. The scale bar = 10 μm.

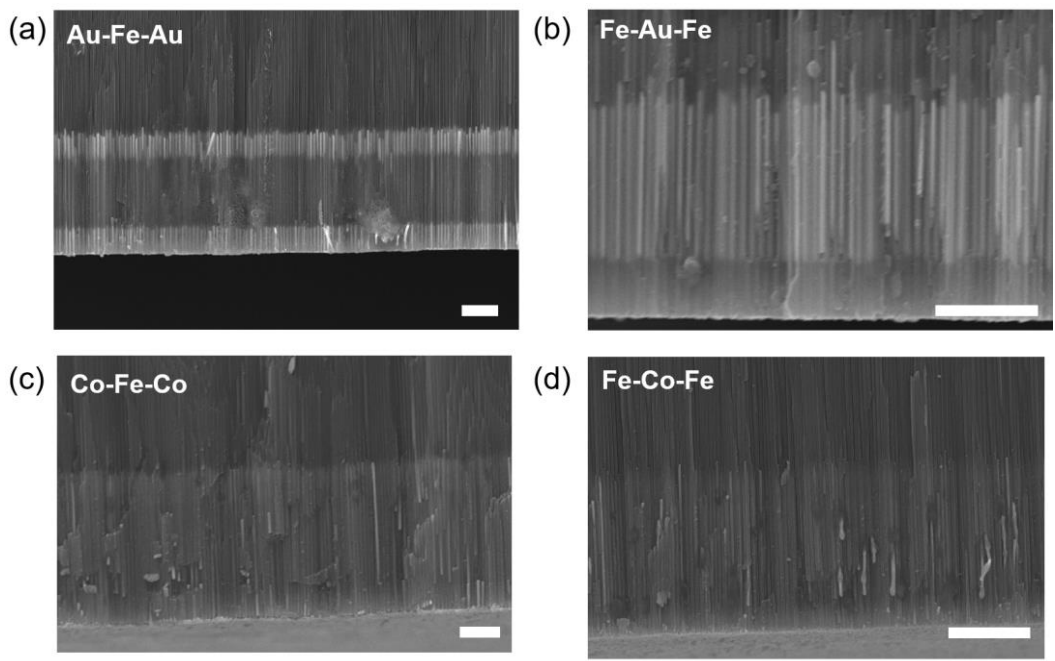

**Fig. S8**. SEM images of the (a) Au–Fe–Au, (b) Fe–Au–Fe, (c) Co–Fe–Co, and (d) Fe–Co–Fe BMN arrays. The scale bar = 5 μm.